\documentclass[%
 reprint,
 superscriptaddress,
 amsmath,amssymb,
 aps,
 pra,
]{revtex4-1}

\usepackage{graphicx}
\usepackage{dcolumn}
\usepackage{bm}
\usepackage{physymb}
\usepackage{calligra}
\usepackage{mathrsfs}
\usepackage{mathtools}
\usepackage{xcolor}

\begin{document}

\title{Crystal Structure Representations for Machine Learning\\Models of Formation Energies}

\author{Felix Faber}%
\affiliation{Institute of Physical Chemistry and National Center for Computational
Design and Discovery of Novel Materials, Department of Chemistry, University of Basel, Switzerland.}
\author{Alexander Lindmaa}%
\affiliation{Department of Physics, Chemistry and Biology, Link\"oping University, SE-581 83 Link\"oping, Sweden.}
\author{O. Anatole von Lilienfeld}%
\email{anatole.vonlilienfeld@unibas.ch}%
\affiliation{Institute of Physical Chemistry and National Center for Computational
Design and Discovery of Novel Materials, Department of Chemistry, University of Basel, Switzerland.}
\affiliation{Argonne Leadership Computing Facility, Argonne National Laboratory, 9700 S. Cass Avenue, Lemont, IL 60439, USA.}
\author{Rickard Armiento}%
\email{rickard.armiento@liu.se}%
\affiliation{Department of Physics, Chemistry and Biology, Link\"oping University, SE-581 83 Link\"oping, Sweden.}

\begin{abstract}
We introduce and evaluate a set of feature vector representations of crystal structures for machine learning (ML) models of formation energies of solids. 
ML models of atomization energies of organic molecules have been successful using a Coulomb matrix representation of the molecule. 
We consider three ways to generalize such representations to periodic systems: 
\emph{(i)} a matrix where each element is related to the Ewald sum of the electrostatic interaction between two different atoms in the unit cell repeated over the lattice; 
\emph{(ii)} an extended Coulomb-like matrix that takes into account a number of neighboring unit cells;
and \emph{(iii)} an {\em ansatz} that mimics the periodicity and the basic features of the elements in the Ewald sum matrix by using a sine function of the crystal coordinates of the atoms. 
The representations are compared for a Laplacian kernel with Manhattan norm, trained to reproduce formation energies using a data set of 3938 crystal structures obtained from the Materials Project. 
For training sets consisting of 3000 crystals, the generalization error in predicting formation energies of new structures corresponds
to (i) $0.49$, (ii) $0.64$, and (iii) $0.37\ \mathrm{eV/atom}$ for the respective representations.
\end{abstract}

\maketitle


\section{\label{sec:intro}Introduction}
First-principles simulations for the prediction of properties of chemical and materials systems have become a standard tool throughout theoretical chemistry, physics and biology. These simulations are based on repeatedly solving numerical approximations to the underlying many-body problem, \emph{i.e.}, the Schr\"odinger equation. However, in the pursuit of ever increasing efficiency, there is a growing interest in side-stepping the physics-based formulation of the problem, and instead exploit big data methodology, \emph{e.g.}, artificial intelligence, evolutionary and machine learning (ML) schemes. If successful, such approaches can lead to an orders-of-magnitude increase in computational efficiency for describing properties of atomistic systems. Such a change would not merely bring incremental progress, but certainly represents a paradigm-shift in terms of enabling the study of systems and problems which hitherto have been completely out of reach.

Recently, one of us has presented an ML scheme that predicts atomization energies of new (out-of-sample) molecules with an accuracy even beyond that of the most common first-principles method, density functional theory \cite{M.Rupp, Hansen.K, von_lilienfeld_first_2013, schutt_how_2014}. Subsequently, it was shown that this approach also works for other properties such as molecular polarizability and frontier orbital eigenvalues \cite{1367-2630-15-9-095003},
transmission coefficients of electron transport in doped nano-ribbon models \cite{PhysRevB.89.235411}, 
and even densities of state within the Andersion impurity model \cite{PhysRevB.90.155136}. 
A critical component of any such ML scheme is the feature vector representation of the data set, also often called the \emph{descriptor} \cite{Fourier_radial_distribution, bartok_representing_2013}; \emph{i.e.}, the mapping of the atomistic description of the system into a form suitable for matrix operations. 
In the cited work~\cite{M.Rupp}, an atom by atom matrix, dubbed \emph{Coulomb matrix}, was used with diagonal and off-diagonal elements representing the potential of the free atom and the inter-atomic Coulomb repulsion between nuclear charges.

The aim of the present study is to consider various ways of adapting this representation to periodic systems, and to compare their performance in ML models of formation energies of solids. 
We have considered three generalizations of the Coulomb matrix idea; \emph{(i)} a matrix where each element is related to the Ewald sum of the electrostatic interaction between two different atoms in the unit cell repeated over the lattice; \emph{(ii)} an extended Coulomb-like matrix that takes into account a number of neighboring unit cells, and \emph{(iii)} a simplified matrix ansatz chosen to mimic the periodicity and basic features of the elements in the Ewald sum matrix using a sine function of atomic coordinates. 

A few studies have already considered representations of periodic systems suitable for ML. Schutt \emph{et al.}~trained an ML model using radial distribution functions to predict the density of states at the Fermi level of solids~\cite{schutt_how_2014}. Meredig \emph{et al.}~used a heuristics based ML model to screen 1.6 M ternary solids for stability~\cite{Wolverton2014}. Ramprasad and co-workers, relying on chemo-structural fingerprints in ML models of formation energies (and other properties) of polymers, obtained scatter plots~\cite{pilania_accelerating_2013} for which visual inspection suggests an error on the order of magnitude of $\sim$0.5 eV. Bartok \emph{et al.}\ used ML to enhance the computer simulation of molecular materials \cite{bartok_machine-learning_2013}. However, to the best of our knowledge no clear example has been presented so far for directly applying ML to reproduce cell-, cohesive-, or formation energies based on an atom by atom matrix based representation of the crystal structure. We present such an application based on the ML methods which have shown good performance for molecules. We find that all representations investigated in this work yield similar accuracy ($0.4-0.6\ \mathrm{eV}/\mathrm{atom}$ at $3000$ crystals), which is less impressive than the accuracy found for atomization energies of molecules ($0.02\ \mathrm{eV}/\mathrm{atom}$ for training sets of similar size). However, our results indicate that if the data set used for training can be further expanded, a machine capable of estimating formation energies at near first-principles accuracy is still within reach. 
We briefly discuss the reason for the disparity in performance between finite and periodic systems, and how this may be resolved.

This paper is organized as follows. 
In Sec.~II we briefly present the kernel ridge regression (KRR) scheme. 
In Sec.~III we introduce our various representations of crystal structures. 
In Sec.~IV some technical details are given regarding the implementation. 
In Sec.~V we analyze the representations and test their performance in machines trained on a data set of formation energies of solids. 
In Sec.~VI we discuss our results, and finally, in Sec.~VII this study is summarized and concluded.

\section{Kernel Ridge Regression}

We present a short summary of the ridge regression method \cite{Ridge_Regression} and KRR \cite{kernel_ridge_regression1fixed, kernel_ridge_regression2}, defining our notation. 
In ridge regression one seeks to approximate an unknown function $y(x)$ where $x$ is a feature-vector representation of some input (e.g., a molecule or crystal system). We start from a set of $n$ data points $\mathbf{y} = (y_1,y_2,...,y_n)^{\mathrm{T}}$ known for a set of feature vectors $\mathbf{x} = (x_1,x_2,...,x_n)^{\mathrm{T}}$, where we use the linear-algebra convention of distinguishing row and column vectors, and $a^{\mathrm{T}}$ is the transpose of $a$.

An approximation $f(x,\boldsymbol{\alpha})$ is constructed as
\begin{eqnarray}
f(x,\boldsymbol{\alpha}) &=& \mathbf{k}(x) \boldsymbol{\alpha} = \sum_{j=1}^n k_j(x)\alpha_j \\
\boldsymbol{\alpha} &=& (\alpha_1,\alpha_2,...,\alpha_n) \\
\mathbf{k}(x) &=& (k_1(x),k_2(x),...,k_n(x))^{\mathrm{T}},
\end{eqnarray}
where $k_j(x)$ is a function \emph{ansatz} of $x$ which is the same for all $j$, and
to be chosen freely, usually assumed to be polynomial. 
The approximation $f(x,\boldsymbol{\alpha})$ is optimized by seeking the 
$\boldsymbol{\alpha}$ that minimizes the Euclidean norm $\left \|\mathbf{y} - \mathbf{k(x)}\boldsymbol{\alpha} \right\|_2$. The solution is given by the normal equation
\begin{equation}
\label{eqn:normalequation}
\boldsymbol{\alpha} =(\mathbf{k(x)}^{\mathrm{T}}\mathbf{k(x)})^{-1}\mathbf{k(x)}^{\mathrm{T}}\mathbf{y}.
\end{equation}
If an \emph{ansatz} for $\mathbf{k}(x)$ with a high degree of freedom is used, there is  considerable risk of overfitting, \emph{i.e.}, one arrives at a model that fits the training data set well, but still performs poorly when predicting $y$ for out-of-sample cases (feature vectors $x$ that are not included in the training set $\mathbf{x}$.) Several techniques are used to address this problem, \emph{e.g.}, cross-validation and regularization. In the case of the latter, Eq.~(\ref{eqn:normalequation}) is modified by inserting an extra term,
\begin{equation}
\label{eq:ridge_regression}
\boldsymbol{\alpha} =(\mathbf{k(x)}^{\mathrm{T}}\mathbf{k(x)} + \lambda
I)^{-1}\mathbf{k(x)}^{\mathrm{T}}\mathbf{y},
\end{equation}
where $\lambda$ is the regularization parameter and $I$ the identity matrix.

Ridge regression is extended into KRR by introducing a map,
$\Phi^K$, from a non-linear space to a linear space that is known as the feature space \cite{muller_introduction_2001,vapnik_nature_1995, schlkopf_learning_2001}.
 The mapping $\Phi^K$
is usually non-trivial, but does not need to be known explicitly since it is sufficient to know the 
inner product between the mapped data, $\langle \Phi^K(x_i),\Phi^K(x_j) \rangle = \sum_l \Phi^K(x_i)_l \Phi^K(x_j)_l = k(x_i,x_j)$ where $k(x_i,x_j)$ is the kernel. This procedure is also known as the \emph{kernel trick} \cite{scolkopf_kernel_trick}. The function $f(x,\boldsymbol{\alpha})$ now takes the form
\begin{equation}
f(x,\boldsymbol{\alpha}) = \sum_{i}^{n}\alpha_i k(x,x_i)
\end{equation}
and the approximation is optimized by seeking the minimum of
\begin{equation}
 \sum_{i}^{n}(y_i - f(x_i,\boldsymbol{\alpha}))^2 + \sum_{i,j}^n\alpha_i k(x_j,x_i) \alpha_j,
\end{equation}
with solution
\begin{equation}
\label{eqn:kkrfit}
 \boldsymbol{\alpha} =(\mathbf{K} + \lambda I)^{-1}\mathbf{y},
\end{equation}
where $\mathbf{K}= k(x_i,x_j)$ is the kernel matrix. 
Different choices of  kernel functions are appropriate for different ML applications. 
Two of the more commonly employed kernel functions are, 
\begin{eqnarray}
\label{eq:kernel}
  \text{Gaussian:}\ k(x_i,x_j) &=& e^{-\left \| x_i-x_j \right\|_{2}^2/(2\sigma^2)} \\
  \text{Laplacian:}\ k(x_i,x_j) &=& e^{-\left \| x_i-x_j \right\|_{1}/\sigma}
\end{eqnarray}
defined as using the Euclidean ($L^2$) and Manhattan ($L^1$) norm, respectively.
Here $\sigma$ represents the kernel width, a measure of locality in the employed training set. 
Without any loss of generality we restrict ourselves to the Laplacian kernel for this study. This choice is motivated by the fact that it has shown good performance in the context of predicting molecular atomization energies \cite{Hansen.K}. However, the Gaussian, or other kernels, could have been used just as well. 

As a relevant benchmark of performance of the ML model with a given representation we use the error in predicting new data outside $\mathbf{x}$, the \emph{generalization error} (GE). There are several ways to estimate the GE. We exclude some subset of all available data $\mathbf{x}$ (and $\mathbf{y}$) from the solution of Eq.~(\ref{eqn:kkrfit}) and then evaluate the mean absolute error (MAE) for the prediction of the excluded data. The subset of $\mathbf{x}$ used in Eq.~(\ref{eqn:kkrfit}) is called the \emph{training set} $\mathbf{x}_\mathrm{train}$, and the remaining input the \emph{test set} $\mathbf{x}_\mathrm{test}$. Hence, we define
\begin{equation}
\label{eq:test_error}
\mathrm{MAE}_{\mathrm{test}} = \frac{1}{m} \sum_{i \in \mathbf{x}_\mathrm{test}} |y_i - f(x_i,\boldsymbol{\alpha})|,
\end{equation} 
where $m$ is the size of the test set. The measure $\mathrm{MAE}_{\mathrm{test}}$ determines how well the machine predicts new data. The number of systems included in $\mathbf{x}_\mathrm{test}$ needs to be sufficiently large for this measure to be an accurate estimate of the GE. Hence, there is a trade-off between using available data as part of $\mathbf{x}_\mathrm{test}$ or $\mathbf{x}_\mathrm{train}$. We alleviate this problem by a statistical validation method known as $k$-fold cross-validation following the recipes in Ref.~\onlinecite{Hansen.K}, combined with random sampling. The $\mathrm{MAE}_{\mathrm{test}}$ is calculated as the mean of the $\mathrm{MAE}_{\mathrm{test}}$ of several independent ML runs. In each run $\mathbf{x}_\mathrm{test}$ is constructed by selecting a specific number of entries randomly out of the full data set.

Similarly, $\mathrm{MAE}_{\mathrm{train}}$ is defined by substituting $\mathbf{x}_{\mathrm{train}}$ for $\mathbf{x}_{\mathrm{test}}$ in Eq.~(\ref{eq:test_error}). This measure determines the precision with which the machine reproduces the data it has been trained on. A high $\mathrm{MAE}_{\mathrm{train}}$ suggests the machine has too few degrees of freedom to describe the data well. A low $\mathrm{MAE}_{\mathrm{train}}$ suggests high flexibility of the fitting function, a potential risk of overfitting, and that it may be possible to reach good transferability with a sufficiently large data set.

\section{Representations}

The feature vector representation used for the input is of central importance for the ML scheme. 
Here, this is the mapping of the positions and identities of all atoms into a vector $x$. 
We expect a well-suited representation to exhibit the following beneficial properties (with examples in parenthesis given for molecules) 
\begin{enumerate}
  \item \textit{Complete, non-degenerate:} 
  $x$ should incorporate all features in the input that are relevant for the underlying problem. (Two different molecules should give different $x$.)  
  \item \textit{Compact, unique:} $x$ should have minimal features that are redundant to the underlying problem. (Two instances of the same molecule, but rotated differently, should give the same $x$.)
 \item \textit{Descriptive:} instances of input that are `close', giving similar $y$, should generally be represented by $x$ that are close, in the sense of a small $\|x_1 - x_2\|$. (Two molecules that are identical except for small differences in the atomic positions, which have similar $y$, should generally have similar $x$.) 
  \item \textit{Simple:} generating the representation for a given input should require as little computational effort as possible. (The set of eigenvalues from solution of the many-body Schr\"odinger equation of the molecule would generally not be a useful $x$.)
\end{enumerate}
A bijective representation is perfectly non-degenerate and unique, \emph{i.e.}, it has a one-to-one correspondence between the input and $x$. In this discussion, the distinction made by `relevant for the underlying problem' is important, because different features of the input may be relevant for different applications. For example, in ML models of atomization energies the chirality of a molecule is not relevant, but in ML models of the optical activity in circular dichroism it is. 

A common method for analysis of the properties of different feature vector representations is principal component analysis (PCA) \cite{PCA}. This method reduces the dimensionality of a data set while still conserving as much information as possible. To generate the PCA, a singular value decomposition \cite{SVD} is performed on $\mathbf{x}$,
\begin{equation}
	\mathbf{x} = \mathbf{U} \boldsymbol{\Sigma} \mathbf{W}^{\mathrm{T}}.
\end{equation}
Submatrices are created from the first $k$ columns extracted from the resulting matrices, $\mathbf{U}_k$ and $\boldsymbol{\Sigma}_k$. The PCA data set of reduced rank $k$ is then defined as
\begin{equation}
	\mathbf{z} = \mathbf{U}_k \boldsymbol{\Sigma}_k.
\end{equation}
This data set is not only useful for visualization. If one suspects that a feature vector representation is not sufficiently compact (in the sense of point 2 in the list of beneficial properties given above in this section), one can try to replace $\mathbf{x}$ with the PCA feature vector representation, $\mathbf{z}$, to extract a representative lower-dimensional part. Note that the PCA representation should only be constructed on the training set. The test set may not be used.

\subsection{Molecular Coulomb matrix}
The present work aims at extending the ML model for molecular properties~\cite{M.Rupp, 1367-2630-15-9-095003, Hansen.K} to properties of crystals. 
We therefore briefly summarize the molecular approach in the following. 
The feature vector representation called the \emph{Coulomb matrix} is a symmetric
atom by atom matrix given in Hartree atomic units,
\begin{eqnarray}
\label{ColumbMatrix}
x_{ij} &=&
  \begin{cases}
    0.5 Z_{i}^{2.4}   &\mbox{if } i = j  \\
    Z_{i}Z_{j} \phi(\|\mathbf{r}_{i}-\mathbf{r}_{j}\|_2) &\mbox{if } i\neq j 
  \end{cases} \\
\label{ColumbMatrix2}
   \phi(r) &=& 1/r
\end{eqnarray}
where $\phi(r)$ is the Coulomb potential, $Z_{i}$ and $\mathbf{r}_{i}$ are the atomic number and position of the $i^{th}$ atom. The non-diagonal elements correspond thus to the pair-wise Coulomb repulsion between the positive atomic cores in the system, and the diagonal elements are chosen by construction as the result of an exponential fit to the potential energy of a free atom. 

While this representation is not bijective, it is non-degenerate in the sense that no two molecules that differ more from each other than being enatiomers will yield the same Coulomb matrix. One molecule, however, can result in several Coulomb matrices due to the various ways of ordering atoms. 
Atom index invariance can be introduced by ordering atom indices according to the norm of each row (or column) \cite{M.Rupp, Hansen.K}, or by using permuted sets of Coulomb matrices \cite{1367-2630-15-9-095003}.
Sorting, however, introduces the issue of differentiability---a potentially desirable property when it comes to the modeling of atomic forces.
Use of the (differentiable) eigenvalue spectrum of the Coulomb matrix yields an atom index invariant descriptor at the expense of losing uniqueness \cite{moussa_comment_2012, rupp_rupp_2012}. 
There is another atom index invariant and differentiable representation, the Fourier series of atomic radial distribution functions, that also preserves uniqueness as well as spatial invariances \cite{Fourier_radial_distribution}. However, this representation has not yet been explored in depth and has therefore not been included in the representations discussed herewithin. 

To benchmark our results for solids, we compare to the performance for the molecular ML model that has been trained and tested on (subsets of) the GDB-database in previous studies \cite{blum_970_2009,fink_virtual_2005,fink_virtual_2007,M.Rupp}. We use a subset of 7165 entries out of the GDB-13 data set. These entries are all the organic molecules in this set with up to 7 atoms of elements C, O, N, and S, and valencies satisfied by hydrogen atoms. The atomic coordinates were relaxed using atomic force fields, and the atomization energies calculated with a higher-order first-principles method, density functional theory \cite{DFT1,DFT2} using the PBE0 hybrid functional \cite{perdew_generalized_1996,ernzerhof_coupling-constant_1997, ernzerhof_assessment_1999}. We call this dataset \emph{QM7} \cite{blum_970_2009,fink_virtual_2005,fink_virtual_2007, ramakrishnan_quantum_2014}. 

When considering ways of extending the Coulomb matrix to periodic systems
one might be tempted to take the Coulomb matrix for just the atoms in the primitive unit cell of the periodic crystal, alongside with the unit cell vectors. However, depending on the choice of primitive unit cell the set of interatomic distances will vary, since distances to neighboring atoms from different unit cells are not accounted for. Hence, such a representation is not unique in the sense that the same crystal can lead to different representations, and therefore is less likely to yield well performing ML models \cite{schutt_how_2014}. 

\subsection{Ewald Sum Matrix}

We now consider a straightforward extension of the Coulomb matrix representation that removes the most obvious dependence on the non-unique set of interatomic distances in the primitive unit cell. We form an atom by atom matrix with one element for each pair of atoms in the primitive unit cell, but now each element is defined to represent the full Coulomb interaction energy corresponding to all infinite repetitions of these two atoms in the lattice. In this way, the elements in the matrix retain essentially the same meaning as in the Coulomb matrix, while the complete infinite repetition of the lattice is taken into account. As such, one can propose the following expression for the matrix elements,
\begin{equation}
\label{eqn:ewaldrep}
x_{ij} = \frac{1}{N} Z_{i}Z_{j} \sum_{k,l,k \neq l} \varphi(\|\mathbf{r}_{k}-\mathbf{r}_{l}\|_2)
\end{equation}
where the sum over $k$ is taken over the atom $i$ in the unit cell and its $N$ closest equivalent atoms, and similarly for $l$ and $j$. The intention is to take $N \to \infty$ to represent the full electrostatic interaction between the infinitely repeated atoms equivalent to atoms $i$ and $j$ in the primitive cell. However, this type of infinite electrostatic sum has well known issues with convergence that have been discussed at length in the field of materials science. One resolution is given by the Ewald sum \cite{Ewald_summation1, Ewald_summation2, martin_electronic_2008}. The central idea is to divide the problematic double sum in Eq.~(\ref{eqn:ewaldrep}) into two rapidly converging sums and one constant,
\begin{equation}
\label{eq:Ewald}
	x_{ij} = x^{(r)}_{ij} + x^{(m)}_{ij} + x^0_{ij},
\end{equation}
where $x^{(r)}_{ij}$ is the short range interaction calculated in real space, $x^{(m)}_{ij}$ the long range interaction calculated in the reciprocal space, and $x^0_{ij}$ is a constant. The division is controlled by a screening length parameter $a$, which influences how rapidly the sums converge. 
The first term is given by
\begin{equation}
\label{eqn:ewaldterm1}
x^{(r)}_{ij} = Z_i Z_j \sum_{\mathbf{L}}  \frac{\mathrm{erfc}(a \|\mathbf{r}_i - \mathbf{r}_j + \mathbf{L}\|_2)}{\|\mathbf{r}_i - \mathbf{r}_j + \mathbf{L}\|_2}\quad (i \neq j),
\end{equation}
where the sum is taken over all lattice vectors $\mathbf{L}$ inside a sphere of a radius set by a cutoff $L_{\mathrm{max}}$. The second term is 
\begin{equation}
\label{eqn:ewaldterm2}
x^{(m)}_{ij} = \frac{Z_i Z_j}{\pi V}  \sum_{\mathbf{G}} \frac{e^{-\|\mathbf{G}\|_2^2/(2 a)^2}}{\|\mathbf{G}\|_2^2} \cos(\mathbf{G} \cdot (\mathbf{r}_i - \mathbf{r}_j))\quad(i \neq j),
\end{equation}
taken over all non-zero reciprocal lattice vectors $\mathbf{G}$ in a sphere of radius set by a cutoff $G_\mathrm{max}$, taken to be large enough for the sum to converge;
and $V$ is the unit cell volume. The last term is 
\begin{equation}
\label{eqn:ewaldterm3}
	x^{0}_{ij} = - (Z_i^2+Z_j^2) \frac{a}{\sqrt{\pi}} - (Z_i+Z_j)^2\frac{\pi}{2 V a^2}\quad(i \neq j),
\end{equation}
where the first term in Eq.~(\ref{eqn:ewaldterm3}) is the Ewald self-terms for the $i$ and $j$ sites and the second term is a correction needed as we use a charged cell, since, in analog to the Coulomb matrix representation, the expressions describe the interaction from the positive atomic cores. The correction makes the total energy be that of a system with a uniform compensating background that makes the system neutral. For the diagonal terms in the matrix ($i$=$j$) we take the Ewald sum interation energy of the lattice of $i$-type atoms, which is given by the same equations Eq.~(\ref{eqn:ewaldterm1}) and (\ref{eqn:ewaldterm2}) but with an extra factor $1/2$ and,
\begin{equation}
	x^{0}_{ii} = - Z_i^2 \frac{a}{\sqrt{\pi}} - Z_i^2\frac{\pi}{2 V a^2}.
\end{equation}
The value of $a$ only affects the rate of convergence in the above sums, not the final value of $x_{ij}$. There are several suggested schemes for how to set it, in our work we take
\begin{equation}
\label{eqn:const}
a = \sqrt{\pi} \left(\frac{0.01 M}{V}\right)^{1/6} 
\end{equation}
where $M$ is the number of atoms in the unit cell.

We refer to Eqs.~(\ref{eqn:ewaldterm1})$-$(\ref{eqn:ewaldterm3}) as the \emph{Ewald sum matrix} representation. Our definitions are chosen to make each element of the matrix be the full Ewald sum of the Coulomb interaction between the sites $i$ and $j$ (or $i$ with itself). We note briefly that the python materials genomics (pytmatgen) open source library \cite{Pymatgen1} contains a similar matrix as an intermediate step in the calculation of the full Ewald sum energy. However, that matrix differ from ours in that it is defined to make the sum over all elements give the total energy in a way that makes individual matrix elements depend on the specific value of the $a$ parameter used. This seems an undesirable feature for using the matrix as a descriptor. (Nevertheless, for the value $a$ in our calculations, we use the same formula as in pymatgen, Eq.~(\ref{eqn:const}).)


\subsection{Extended Coulomb-like Matrix}\label{sec:ecm}

Another way to generalize the Coulomb matrix to periodic systems is to extend the size of the representation matrix. Let each element be the electrostatic interaction between one of the $M$ atoms in the unit cell and one of the atoms in the $N$ closest unit cells, giving an $M$ by $N \cdot M$ matrix representation on the regular Coulomb matrix form of Eq.~(\ref{ColumbMatrix}), with $1\leq i \leq M$, $1 \leq j \leq N\cdot M$. To completely avoid the dependence on the chosen primitive unit cell, one would like to take $N \to \infty$, \emph{i.e.,} an infinitely large representation matrix. This can be avoided if the long-range electrostatic interaction is replaced with a more rapidly decaying interaction. Here, we chose $\tilde \varphi(r) = e^{-r}$. In this way the elements of the matrix quickly drop to zero, and the representation matrix can be cut off at a finite dimension that is taken to be sufficiently large for all systems in the data set. We refer to this as the \emph{extended Coulomb matrix} representation. One benefit of this representation over the Ewald sum matrix is that it is more straightforward to evaluate (one can simply iterate over $N$ copies of the unit cell). 

\subsection{Sine Matrix}
Yet another representation can be constructed by further extending the idea of reducing the computational effort by replacing the long-range electrostatic interaction by a simpler expression. We start from the $M$ by $M$ Ewald sum matrix in Eq.~(\ref{eqn:ewaldrep}), but substitute the whole sums of the electrostatic interaction with an arbitrarily chosen two-point potential that is intended to share the same basic properties as such a sum, giving
\begin{equation}
	\label{eq:sinusoidal}
	x_{ij} =
	  \begin{cases}
		0.5 Z_{i}^{2.4}   &\mbox{if } i = j  \\
		Z_{i}Z_{j} \tilde \Phi(\mathbf{r}_i,\mathbf{r}_j)
		&\mbox{if } i\neq j
	  \end{cases}
\end{equation}
where $\mathbf{r}_{l}$ is the position of the $l^\mathrm{th}$ atom in the unit cell.

Consider two non-equivalent atoms in the unit cell, $A$ and $B$. The Coulomb sum contribution due to the infinitely repeated grid of $A$ and $B$ atoms can be thought of as a potential field which is a function of the position of the atom $A$. Three important properties of this field are: \emph{(i)} the expression as function of each atomic coordinate is periodic with respect to the crystal lattice; \emph{(ii)} the contribution from two equivalent atoms in neighboring cells should be the same; \emph{(iii)} the potential should approach infinity when $A$ takes the same position as $B$. The conclusion is that the potential needs to be symmetric with respect to the lattice vectors. 

\begin{figure}
	\centering
        \includegraphics[width=\linewidth]{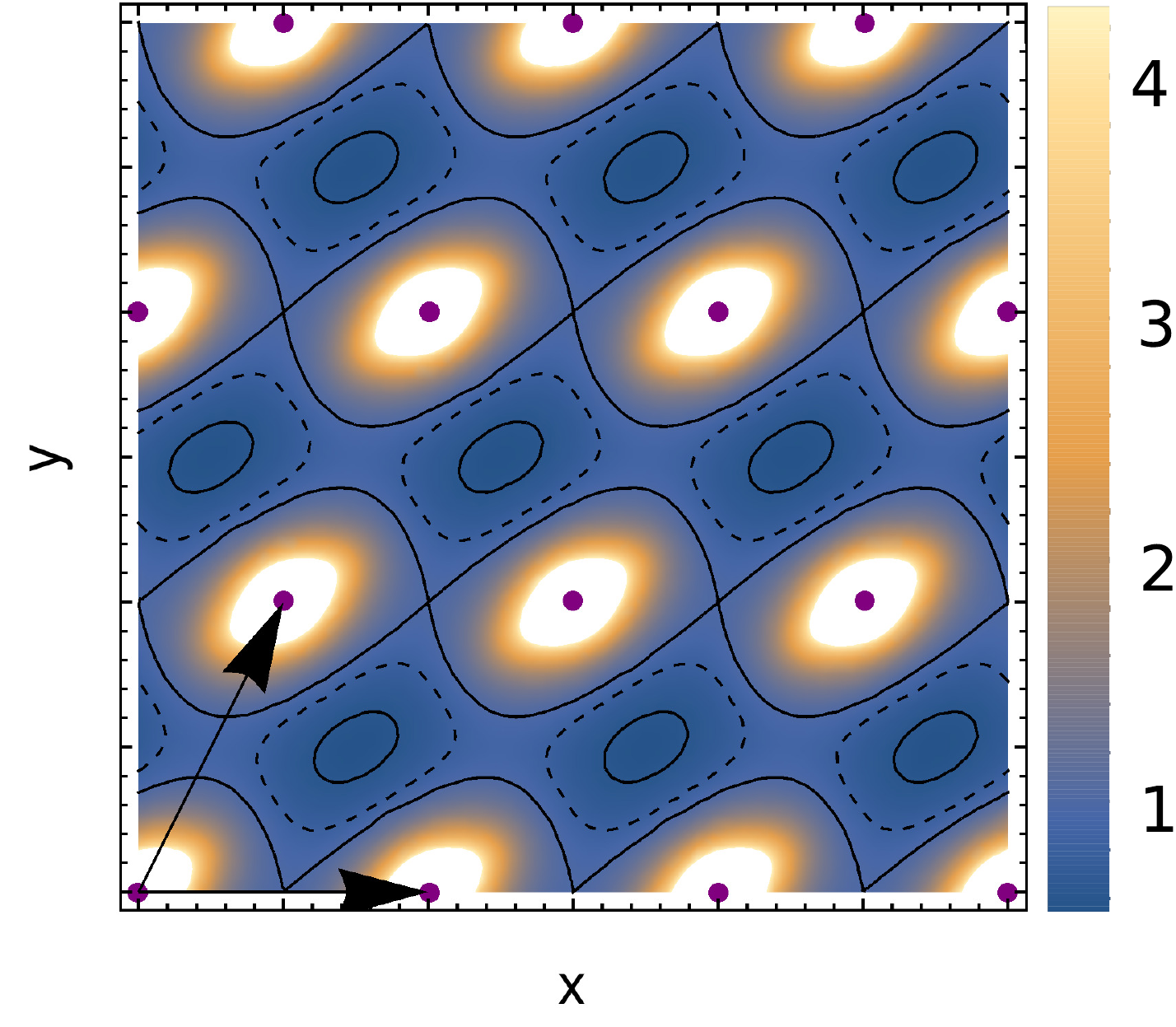}
	\caption{\label{fig:2DSinusodial}
        An illustration of $\tilde \Phi(\mathbf{r}_1,\mathbf{r}_2)$ used in the sine matrix representation, Eq.~(\ref{eq:sinusoidal2}), for a two-dimensional crystal lattice in a primitive unit cell shown by arrows. The figure shows the magnitude of our constructed `interaction' between one atom at $\mathbf{r}_1 = (x,y)$ and another fixed at the origin $\mathbf{r}_2=0$ (the latter shown along with its infinite repetitions as solid purple dots.) The interaction is periodic across the unit cells and grows to infinity as $\mathbf{r}_1$ approach any repetition of the atom at the origin.        
        }
\end{figure}

A possible choice for $\tilde \Phi(\mathbf{r}_1,\mathbf{r}_2)$ that fulfills these requirements is 
\begin{equation}
 \label{eq:sinusoidal2}
	\tilde \Phi(\mathbf{r}_1,\mathbf{r}_2) =\|\mathbf{B} \cdot \sum_{\mathclap{k=\{x,y,z\}}} \hat e_k \sin^2\boldsymbol{[}\pi \hat e_k \mathbf{B^{-1}} \cdot (\mathbf{r}_1 - \mathbf{r}_2)\boldsymbol{]} \|_{2}^{-1}
\end{equation}
where $\hat e_x, \hat e_y, \hat e_z$ are the coordinate unit vectors and $\mathbf{B}$ is the matrix formed by the basis vectors of the lattice. The product inside the sine function thus gives the vector between the two sites expressed in crystal lattice coordinates, which gives the right periodicity in $\mathbf{r}_1$ and $\mathbf{r}_2$. We call Eqs.~(\ref{eq:sinusoidal})$-$(\ref{eq:sinusoidal2}) the \emph{sine matrix} representation. The benefit of this representation over the others suggested in this work is that Eq.~(\ref{eq:sinusoidal}) is a completely straightforward $M$ by $M$ matrix that only depends on the positions of the atoms in a single unit cell. Hence, the computational load of this representation is minimal. Figure~\ref{fig:2DSinusodial} shows $\tilde \Phi$ for a two-dimensional lattice. 

Note that we do not have a proof for the completeness or uniqueness of these representations. They are merely constructed as sensible extensions of the Coulomb matrix idea.

\section{Implementation Details and Data sets} \label{sec:techdetails}

We have implemented the KRR ML scheme using a Laplacian kernel both for the original set of molecules from Ref.~\onlinecite{M.Rupp} with the Coulomb matrix representation, and for the discussed representations using a data set of cystal structures. We use the Python programming language, including numpy \cite{van_der_walt_numpy_2011}, scipy \cite{scipyweb} and pytmatgen \cite{Pymatgen1}.

\begin{figure}
\centering
\includegraphics[width=\linewidth]{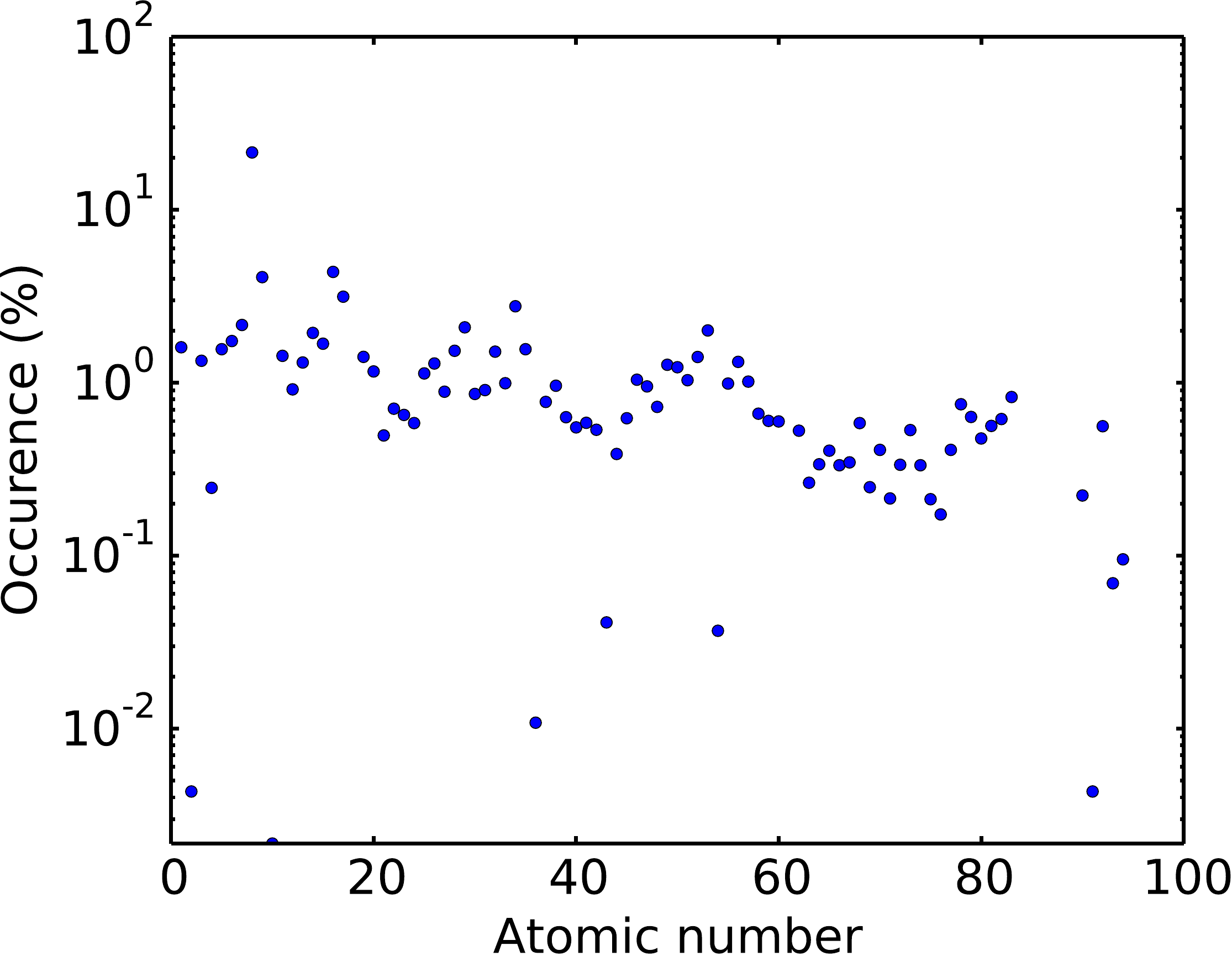}
\caption{
\label{fig:ElementOccurence_Unsorted}
The frequency  of occurrence of various elements in the 3938 systems in the MP data set. The distribution is not uniform, but rather expected to reflect the occurrence of elements in published materials. The four most common elements are O, Si, Cu, and S.}
\end{figure}

Our data set of formation energies of periodic solids was obtained from the \emph{Materials Project} (MP) database \cite{materals_project1}. We extracted $3938$ systems from the MP without obvious order. Since the MP database is derived from the ICSD \cite{bergerhoff_inorganic_1983, belsky_new_2002}, the distribution of elements in the extracted systems roughly match their occurrence in published materials in the literature. This distribution is shown in Fig.~\ref{fig:ElementOccurence_Unsorted}. While this may be interpreted as a bias in the data set, one can also see it as representative for the intended application of ML in predicting properties of materials out of available material databases based on published materials.

The machine requires values for the regularization parameter $\lambda$ and kernel width $\sigma$. We follow Hansen \emph{et al.} \cite{Hansen.K}; optimal values were identified by calculating $\mathrm{MAE}_{\mathrm{test}}$ for a training set size of 3000 for pairs of $\lambda$ and $\sigma$ values on a two-dimensional logarithmic grid, using a spacing factor of 2 for $\sigma$ and 10 for $\lambda$. (This method works well when the model has a small number of hyper parameters, but needs to be replaced with more sophisticated methods for ML models with a larger set of parameters since the time it takes to find the optimum is of $\mathcal{O}(x^p)$ where $p$ is the number of hyper parameters.) The optimal values found are shown in Table~\ref{tab:HyperParameterValues}. The $\mathrm{MAE}_{\mathrm{test}}$ for the different representations is not very sensitive to these parameters, \emph{i.e.}, the regions around the minimums in the generated grids were relatively flat.

\begin{table}
\caption{\label{tab:HyperParameterValues}
Optimal values of parameters $\lambda$ and $\sigma$ for the machines using the different feature vector representations in this work, and for a training set consisting of 3000 crystals drawn at random from the Materials Project. }
\begin{ruledtabular}
\begin{tabular}{@{\ }l c c@{\ }}
  Representation & $\sigma $    &    $\lambda$ \\
\hline
  Ewald sum matrix 	& $1.0 \cdot 10^5$    &    $10^{-4}$ 	\\
  Generalized Coulomb matrix & $8.0 \cdot 10^4$    &    $10^{-3}$ 	\\
  Sine matrix 	& $4.0 \cdot 10^4$    &    $10^{-4}$ 	\\
  Coulomb matrix for QM7 molecules & $2.5 \cdot 10^3$  &    $10^{-6}$ 	\\
\end{tabular}
\end{ruledtabular}
\end{table}

\section{Results}
\label{Results1}

\begin{figure}
	\centering
	\includegraphics[width=\linewidth]{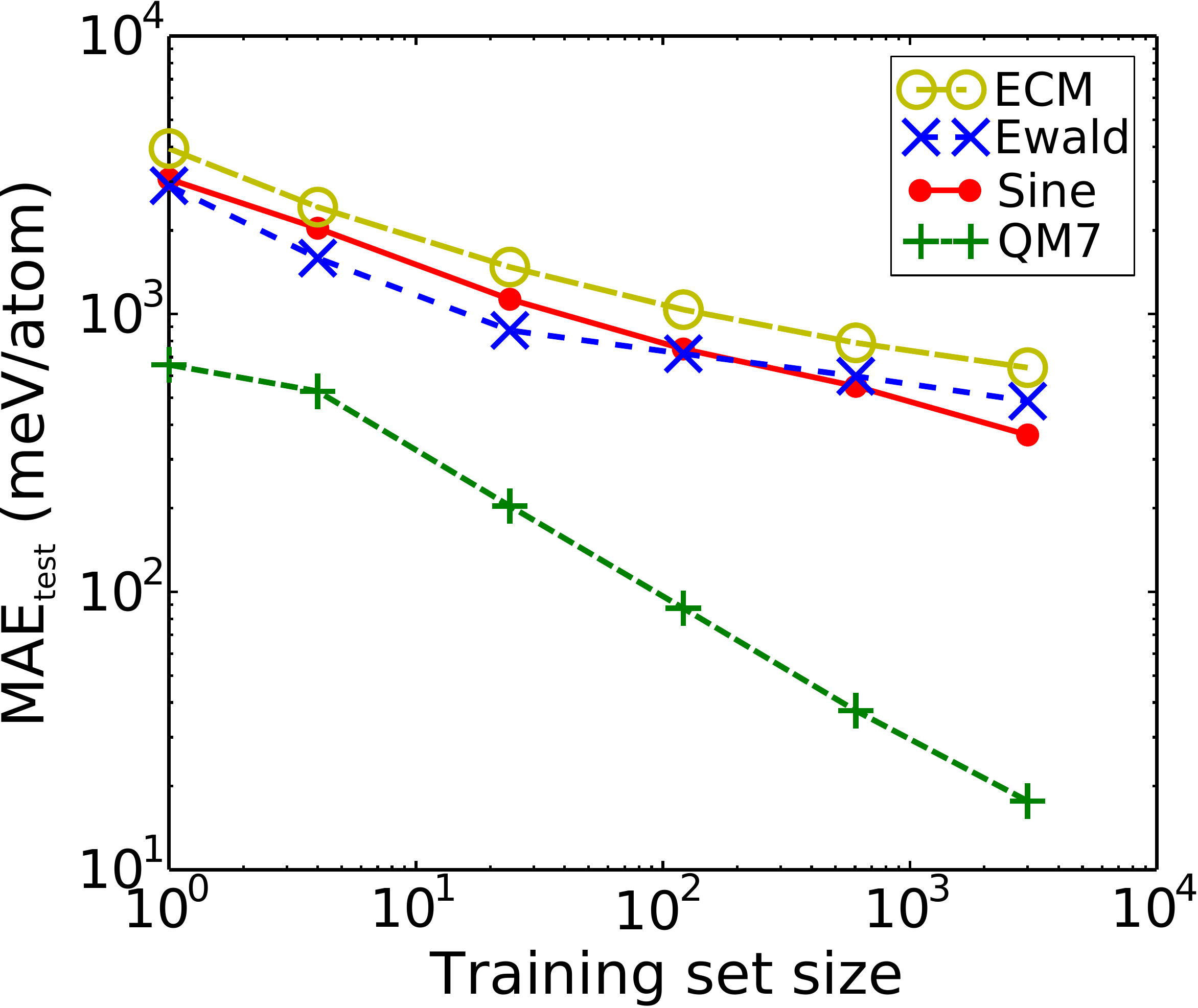}
	\caption{
	\label{fig:ErrorGraph}
The mean absolute GE, Eq.~(\ref{eq:test_error}), versus training set size for the different representations considered in this work. Shown are $\mathrm{MAE}_\mathrm{test}$ for predicting formation energies of crystals in the MP data set: (Sine) Eqs.~(\ref{eq:sinusoidal})$-$(\ref{eq:sinusoidal2}); (Ewald)  Eqs.~(\ref{eqn:ewaldterm1})$-$(\ref{eqn:ewaldterm3}); and (GCM) described in Sec.~\ref{sec:ecm}. For comparison we also include (QM7), $\mathrm{MAE}_\mathrm{test}$ for atomization energies of molecules in the QM7 data set using a regular Coulomb matrix, Eqs.~(\ref{ColumbMatrix})$-$(\ref{ColumbMatrix2}).}
\end{figure}

\begin{figure}
	\centering
	\includegraphics[width=\linewidth]{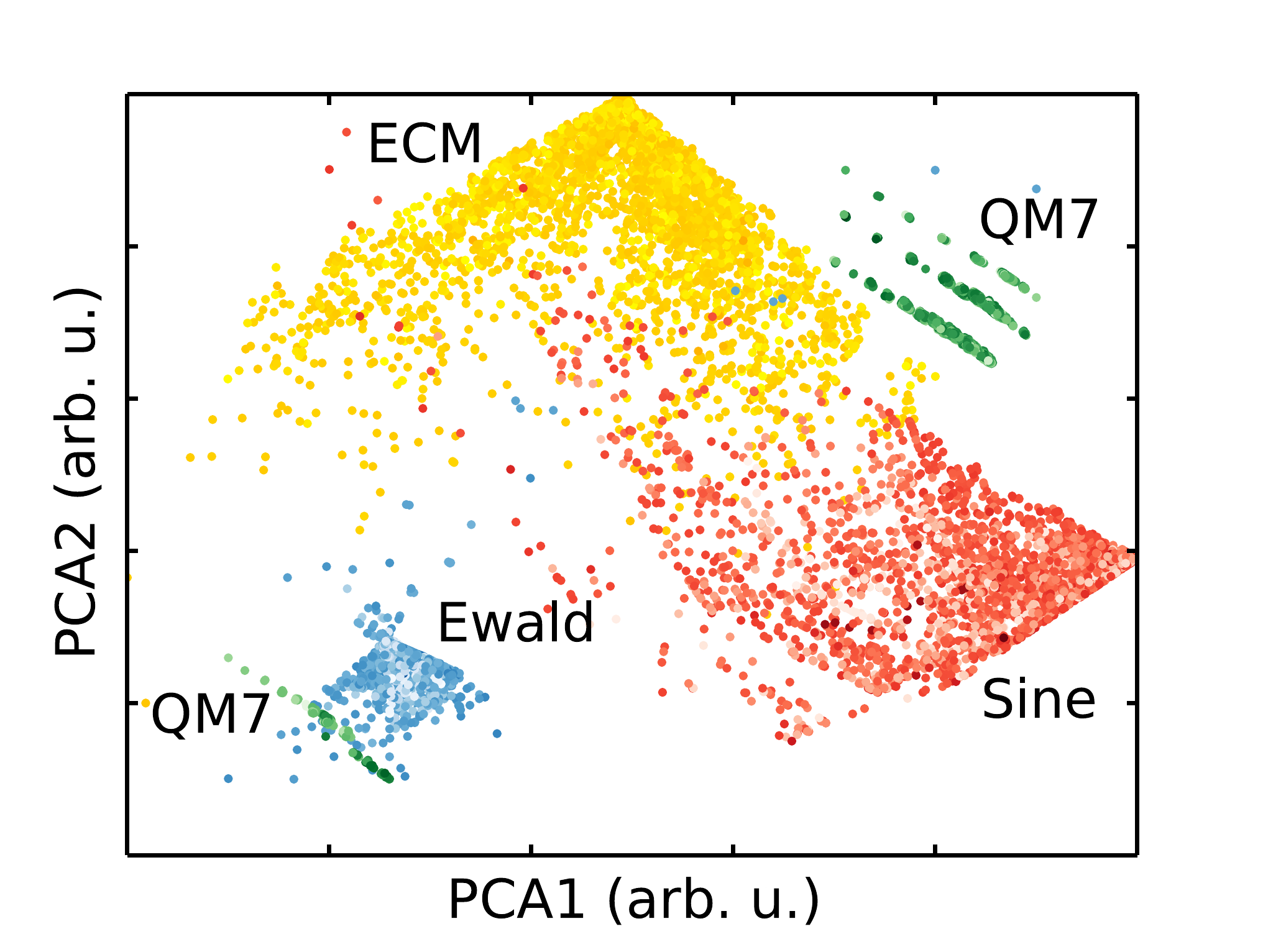}
	\caption{
	\label{fig:PCA}
	A two-dimensional PCA of the data sets used in this work: (QM7, green) the Coulomb matrix representation of the QM7 data set of molecules; (Sine, red) the sine matrix representation using the MP data set; (Ewald, blue) the Ewald sum matrix representation using the MP data set; (ECM, yellow) The extended Coulomb matrix data set using the MP data set. The PCA of the representations of the MP data set differ substantially from the PCA of the Coulomb matrix of QM7. The shading of the points represent the target value energy, showing that there is no clear energy-PCA pattern.}
\end{figure}

The performance of the representations of periodic systems studied in this work are meant to be considered in the context of the excellent performance of the machine for molecules presented in Refs.~\onlinecite{M.Rupp, Hansen.K}. To make this comparison clear we have reproduced the GE of their scheme with our implementation and the QM7 data set. The results are shown in Fig.~\ref{fig:ErrorGraph}. In Fig.~\ref{fig:PCA} a two-dimensional PCA of the QM7 set using the Coulomb matrix representation is shown. 
Already with a training set size of a couple of hundred atoms, the MAE approaches the accuracy of DFT with the least computationally expensive, semi-local, functionals. At a training set size of 500, we arrive at a $\mathrm{MAE}_{\mathrm{test}}$ of around $0.07\ \mathrm{eV}/\mathrm{atom}$, whereas DFT with the PBE functional has an MAE for the atomization energy of small molecules of ca $0.15\ \mathrm{eV}/\mathrm{atom}$ \cite{perdew_generalized_1996}. As demonstrated by Rupp \emph{et al.}\ \cite{M.Rupp}, the precision of the predictions of the machine keeps improving with increased size of the training set, and at 3000 structures one finds an GE error below $0.02\ \mathrm{eV}/\mathrm{atom}$.

We now turn to the results of applying ML to periodic crystals using the feature vector representations in this work, shown in Fig.~\ref{fig:ErrorGraph}. The performance of all our representations are similar, but their accuracy is inferior to what we see for molecules. All the representations studied improve greatly with increased training set size. The $\mathrm{MAE}_{\mathrm{test}}$ for the representations at a training set size of $3000$ are $0.49\ \mathrm{eV}/\mathrm{atom}$ for the Ewald sum matrix, $0.64\ \mathrm{eV/atom}$ for the extended Coulomb matrix, and $0.37\ \mathrm{eV/atom}$ for the sine matrix representation, \emph{i.e.}, an order of magnitude worse than we find for the molecules. Furthermore, we find that $\mathrm{MAE}_\mathrm{train}$ for all the representations are insignificant ($< 5 \cdot 10^{-3}$) even at a training set size of 3000.

Figure \ref{fig:PCA} compares the two-dimensional PCA of the MP data set to the QM7 one for the different representations. The QM7 PCA is localized to a set of small clusters. The MP data for the sine and the extended Coulomb matrix  representations appear more uniformly spread out. The Ewald sum matrix PCA has a central cluster but also a few data points far removed from this cluster. 

\section{Discussion}
The results in Fig.~\ref{fig:ErrorGraph} suggest that the sine matrix representation is slightly better than the other representations in this work, in that it reaches a lower GE and has better development of the GE with training set size. However, the performance of the different methods are roughly similar, indicating that the selection between the feature vector representations presented in this work for a given application may not have major impact on the GE. This further strengthens the case of the sine matrix representation, since it is the representation that requires the least computational expense. It is interesting to note that while the PCAs of the sine and extended Coulomb matrix bear strong resemblance, the Ewald sum PCA is distinctly different with a strong clustering  and a few outlier points. Furthermore, the GE of all representations systematically decreases with increasing training set size. These results suggest the GE could be reduced even further if only more extensive data set for training were available.

When comparing the molecular results with the ones for solids, it is important to keep in mind that the diversity and composition of the MP data set differs substantially from that of QM7. 
In particular, only very few element types are present in the molecular data set
consisting of atoms with very finite size, no molecule has more than seven atoms (not counting
hydrogens).
This is illustrated by the PCA in  Fig.~\ref{fig:PCA}, where the QM7 data is collected in a few tight clusters. 
Arguably, the chemical space of QM7 is significantly smaller than the one we use to evaluate the representations for solids since the MP data set contains ten times more elements than QM7. 

Hence, the central conclusion of the present work is that there are three areas where the situation of ML for periodic systems can be improved: \emph{(i)} our methods should be used with even larger data sets to confirm that the GE can be brought down to levels where it is useful for applications, \emph{(ii)} further improved representations may be helpful if they more efficiently can represent the degrees of freedom offered by periodic crystals. Such representations may reduce the need for larger training sets; and \emph{(iii)} if a  way of generating a data set over a restricted chemical space can be devised which is as compact as QM7, we may reach more promising ML performance for periodic systems. We are presently working in all of these directions.

\section{Summary and Conclusions}
We have investigated the performance of several crystal structure representations for ML models of formation energies of solids. Our work is a natural generalization of an ML scheme previously shown to be successful for the atomization energy of molecules. We have compared three different representations, and found that a sine matrix  which simulates the features of an infinite Coulomb sum is both most efficient, and gives the smallest GE error. While the performance of all the methods may at first seem disappointing when compared to the small GE confirmed for molecules, we can explain this discrepancy to be due to data sets which are too small to cover the hugely diverse compositional and structural space with sufficient density. As such, the full potential of ML for periodic systems still has to be demonstrated. 
The improvement of the $\mathrm{MAE}_\mathrm{test}$ with training set size suggests, however, that the methods presented here can lead to accurate machine models if only trained on larger or more restricted data sub sets. Hence, our results offer promising indications that sufficiently accurate and transferable ML models of energies of periodic systems can be realized.

\section{Acknowledgments}
R.A. acknowledges funds provided by the Swedish Research Council Grant No. 621-2011-4249 and the Linnaeus Environment at Link\"oping on Nanoscale Functional Materials (LiLi-NFM) funded by VR. Calculations have been performed at the Swedish National Infrastructure for Computing (SNIC).
O.A.v.L. acknowledges funding from the Swiss National Science foundation (No.~PP00P2\_138932).
This research used resources of the Argonne Leadership Computing Facility at Argonne National Laboratory,
which is supported by the Office of Science of the U.S.~DOE under contract DE-AC02-06CH11357.
This material is based upon work supported by the Air Force Office of Scientific Research, Air Force Material Command, USAF under Award No. FA9550-15-1-0026.
   
\bibliographystyle{apsrev4-1}
\bibliography{allrefs}

\end{document}